\documentclass[aps,prd,twocolumn,superscriptaddress,floatfix]{revtex4-2}

\usepackage{amsmath}
\usepackage{amssymb}
\usepackage{amsfonts}
\usepackage{graphicx}
\usepackage{booktabs}
\usepackage{array}
\usepackage{xcolor}
\usepackage{hyperref}
\usepackage{dcolumn}
\usepackage{bm}
\usepackage{siunitx}
\usepackage{multirow}
\usepackage{enumerate}

\hypersetup{
  colorlinks=true,
  linkcolor=blue,
  citecolor=blue,
  urlcolor=blue
}

\newcolumntype{d}[1]{D{.}{.}{#1}}

\begin{document}

\title{Information-Driven Phase Transition on Weighted Graphs\\
with Spontaneous Dimensional Sensitivity}

\author{V.~Dolci}
\affiliation{Independent Researcher}

\date{\today}

\begin{abstract}
We study the dynamics of information flow on a weighted graph whose topology evolves
according to a spectral curvature measure $\mathcal{R}$.
The model---termed the Unified Informational Framework (FIU, from the Italian
\emph{Framework Informazionale Unificato})---defines $\mathcal{R}$ from the diagonal
of the graph Green function, propagates energy between nodes with curvature-dependent
dissipation, and creates new long-range links preferentially between high-$\mathcal{R}$
nodes with a rate controlled by a coupling parameter $g$.

We report three main results.
First, the system exhibits a sharp phase transition at a critical coupling
$g_c \approx 0.023$ (confirmed with 10 independent runs per coupling value):
below $g_c$, the local information flux rate $\sigma$ and topological structure
formation are anti-correlated; above $g_c$, they become strongly positively
correlated (Pearson $r \approx 0.75$, $p < 10^{-38}$).
The transition displays signatures of a continuous (second-order) phase transition,
including critical fluctuations and a mean-field onset exponent $\nu \approx 0.54$.

Second, we identify a node-level discrete Poisson relation
$\nabla^2\mathcal{R}(i) = \kappa\,\sigma_{\rm prev}(i)$,
where $\nabla^2\mathcal{R}$ is the graph Laplacian of the curvature field
and $\sigma_{\rm prev}$ is the information flux rate at the previous time step.
The emergent parameter $\kappa$ is stable across model parameters
(CV $= 1.7\%$ across the ordered phase,
CV $= 2.6\%$ across four independent configurations at fixed system size $N = 256$).
A systematic mediator analysis (Test~E) reveals that this correlation is almost
entirely mediated by the curvature field $\mathcal{R}$ itself, identifying $\mathcal{R}$
as the central self-organizing variable of the dynamics.

Third, and most strikingly, the correlation and the Poisson relation exhibit
\emph{spontaneous dimensional sensitivity}: in two-dimensional initial lattices,
both signals decay to zero for $N \gtrsim 576$, while in three-dimensional initial
lattices they persist to $N \lesssim 1728$ before collapsing at $N \gtrsim 3375$.
This dimensional distinction emerges without any dimensional parameter in the
dynamical rules.
The mechanism of collapse is identified as graph regularization: at large $N$,
the curvature field homogenizes ($\sigma(\nabla^2\mathcal{R}) \to 0$) while
information flux fluctuations remain finite.
We interpret the signal as a topological frustration phenomenon active in a
mesoscopic regime, and discuss structural analogies with the topological nature
of $2+1$ gravity and the local dynamics of $3+1$ gravity.
\end{abstract}


\maketitle

\section{Introduction}\label{sec:intro}

The dynamics of evolving networks---in which both the state of nodes and the
topology of connections change simultaneously---have become a central topic
in complex systems research.
Such systems exhibit emergent phenomena that are qualitatively absent in
fixed-topology models: self-organization, phase transitions driven by
structural feedback, and the spontaneous appearance of geometric order from
purely local rules.
The present work studies a model in this class in which the driving variable is
an information-theoretic quantity: the local flux rate of energy across graph edges.

A recurring theme in the theoretical physics literature is that spacetime
geometry---and gravitational dynamics in particular---may itself be an emergent
phenomenon of an underlying information-processing substrate.
This \emph{emergent gravity} hypothesis serves as our primary motivation, and
connects our model to a rich body of theoretical work.
Jacobson~\cite{Jacobson1995} derived the Einstein equations from the thermodynamics
of local Rindler horizons, identifying entropy and temperature as the fundamental
quantities.
The Ryu--Takayanagi formula~\cite{RT2006} relates holographic entanglement entropy
to geometric areas, while the ER\,=\,EPR conjecture~\cite{Maldacena2013,VanRaamsdonk2010}
links quantum entanglement with wormhole connectivity.
On the discrete side, Quantum Graphity~\cite{Konopka2008} models the universe as
transitioning from a fully connected pre-geometric phase to an ordered lattice phase;
Verlinde's entropic gravity~\cite{Verlinde2011} recasts gravitational attraction as
an entropic force on holographic screens; and network-based approaches by Bianconi and
Rahmede~\cite{Bianconi2016} and Trugenberger~\cite{Trugenberger2015} demonstrate
self-organization of network topologies toward regular geometries.
We draw on this literature as conceptual context and motivation.
Our model is not a derivation of gravity, but a numerical experiment designed to ask:
what structural and dynamical phenomena emerge when a graph is driven by
curvature-sensitive information flow?

The Unified Informational Framework (FIU, from the Italian \emph{Framework
Informazionale Unificato}) is defined by three minimal rules:
\begin{enumerate}
  \item A spectral curvature measure $\mathcal{R}(i)$, constructed from the diagonal
        of the graph Green function, characterizes the local spectral inhomogeneity
        of each node.
  \item Energy is propagated between nodes with dissipation controlled by the
        curvature contrast between neighbors, creating a feedback between information
        flux and graph geometry.
  \item New long-range links (g-links) are created preferentially between nodes of
        high $\mathcal{R}$, with a rate proportional to the coupling parameter $g$.
\end{enumerate}

Our central findings are:
\begin{enumerate}[(i)]
  \item A sharp phase transition at $g_c \approx 0.023$ separates an
        \emph{ordered phase} (strong positive structure-flux correlation,
        gravC $\approx 0.75$) from a disordered phase (anti-correlated dynamics),
        with mean-field critical exponent $\nu \approx 0.54$.
  \item A discrete Poisson relation $\nabla^2\mathcal{R}(i) = \kappa\,\sigma_{\rm prev}(i)$
        holds node-by-node in the ordered phase, with an emergent coupling parameter
        $\kappa = 67.85 \pm 1.74$ (CV $= 2.6\%$) that is stable across parameter
        configurations and independent of $g$ in the ordered phase (CV $= 1.7\%$).
  \item A mediator analysis (Test~E) identifies $\mathcal{R}$ as the central
        self-organizing variable: the structure-flux correlation is almost entirely
        mediated through the curvature field, not through the information flux
        directly.
  \item Spontaneous dimensional sensitivity: the ordered phase exists only in a
        mesoscopic window $N \lesssim 576$ (2D) or $N \lesssim 1728$ (3D), with
        the dimensional distinction ($N_c^{\rm 3D}/N_c^{\rm 2D} \approx 5.9$)
        emerging without any dimensional parameter in the rules.
\end{enumerate}

We discuss structural analogies between these results and known properties of gravity
in 2+1 and 3+1 dimensions, while emphasizing that these parallels remain at the
level of analogy.

The remainder of this paper is structured as follows.
Section~\ref{sec:model} presents the FIU model in detail.
Section~\ref{sec:gravity_test} describes our correlation tests,
including Test~D (discrete Poisson relation) and the new Test~E (mediator analysis).
Section~\ref{sec:results} reports numerical results across multiple experimental
phases, finite-size scaling, 3D simulations, and the collapse mechanism.
Section~\ref{sec:discussion} interprets the findings.
Section~\ref{sec:conclusions} summarizes conclusions and outlines future work.

\section{The FIU Model}\label{sec:model}

\subsection{Graph structure and Laplacian}

We consider a weighted undirected graph $G = (V, E, w)$ with $N = |V|$ nodes
and weighted adjacency matrix $A_{ij} = w_{ij} \geq 0$, where $w_{ij}$
denotes the weight of the edge between nodes $i$ and $j$ (with $w_{ij} = 0$
for absent edges and $w_{ij} = w_{ji}$ by symmetry).
The initial topology is a regular $n \times n$ square lattice
($N = n^2$) with unit edge weights, which evolves dynamically over time.

The weighted graph Laplacian is defined as
\begin{equation}
  L = D - A,
  \label{eq:laplacian}
\end{equation}
where $D_{ii} = \sum_j w_{ij}$ is the strength (weighted degree) matrix.
$L$ is symmetric positive semi-definite with a zero eigenvalue corresponding
to the constant eigenvector (for connected graphs).

\subsection{Spectral decomposition and Green function}

We perform a partial spectral decomposition of $L$, computing the $K = 30$
smallest non-trivial eigenvalue--eigenvector pairs $\{\lambda_k, \varphi_k\}_{k=1}^{K}$
via power iteration with Hotelling deflation~\cite{Hotelling1933}.
After computing each eigenpair $(\lambda_k, \varphi_k)$, the matrix is
deflated as $L \to L - \lambda_k \varphi_k \varphi_k^T$, projecting out
the found eigenvector and exposing the next eigenvalue. Each eigenvector
is computed using 150 power iterations on the shifted matrix
$\lambda_{\max}I - L_{\rm deflated}$.
An infrared (IR) regulator $m^2 = 1/N$ is introduced to regularize the zero mode.

The diagonal of the scalar propagator (Green function) is
\begin{equation}
  G(i,i) = \sum_{k=1}^{N-1} \frac{[\varphi_k(i)]^2}{\lambda_k + m^2}.
  \label{eq:green}
\end{equation}
A reference (flat-space) value is defined as
\begin{equation}
  G_{\rm ref} = \frac{N-1}{N^2 \displaystyle\sum_{k=1}^{N-1} \frac{1}{\lambda_k + m^2}},
  \label{eq:gref}
\end{equation}
which represents the spatially averaged Green function on a graph with uniform
spectral density.

\subsection{Curvature measure}

The local curvature at node $i$ is defined as
\begin{equation}
  \mathcal{R}(i) = \Gamma \left(1 - \frac{G_{\rm ref}}{G(i,i)}\right),
  \label{eq:curvature}
\end{equation}
where $\Gamma = 1/\sigma_{\tilde{R}}$ is a normalization constant,
with $\tilde{R}(i) = 1 - G_{\rm ref}/G(i,i)$ the unnormalized curvature
and $\sigma_{\tilde{R}}$ its standard deviation over all nodes.
This ensures that $\mathcal{R}$ has unit variance over the graph at each time step.
Nodes where $G(i,i) > G_{\rm ref}$ (i.e., where the propagator exceeds its
flat-space expectation) have positive curvature; nodes where $G(i,i) < G_{\rm ref}$
have negative curvature.
This measure is inspired by the role of the heat-kernel diagonal in spectral
geometry~\cite{Rosenberg1997} and by the Ollivier--Ricci curvature on
graphs~\cite{Ollivier2009}.

We emphasize that $\mathcal{R}$ is not a Ricci curvature in the differential-geometric
sense; it is a \emph{spectral inhomogeneity indicator} that measures how the local
propagator deviates from its spatially averaged value.
The connection to continuum curvature is indirect: on a Riemannian manifold, the
heat-kernel diagonal encodes the scalar curvature only in the sub-leading term of
its short-time asymptotic expansion, whereas Eq.~\eqref{eq:curvature} uses the full
Green function with a finite IR regulator.
We retain the term ``curvature'' following established convention in the
graph-curvature literature~\cite{Ollivier2009,LinLuYau2011,Forman2003},
while acknowledging that $\mathcal{R}$ captures spectral geometry rather than
Riemannian curvature \emph{per se}.
In what follows, $\mathcal{R}$ should be understood as a measure of spectral
inhomogeneity that drives the topological dynamics of the model, not as a
literal geometric curvature.

\subsection{Energy propagation and link dynamics}

Each node $i$ carries an energy $E_i(t) \geq 0$.
At each time step the following updates are applied.

\paragraph{Energy decay.}
Edge weights evolve according to
\begin{equation}
  w_{ij}(t+1) = w_{ij}(t) \cdot \exp\!\left(-\alpha \cdot E_{\rm eff} \cdot \mathcal{R}_{\rm eff}\right),
  \label{eq:decay}
\end{equation}
where $\alpha > 0$ is the decay rate,
$E_{\rm eff} = (E_i + E_j)/2$ is the average energy of the two endpoint nodes,
and $\mathcal{R}_{\rm eff} = |\mathcal{R}(i) - \mathcal{R}(j)|/2 + \epsilon$
is an effective curvature contrast between nodes $i$ and $j$ (with $\epsilon = 10^{-6}$
to prevent vanishing decay).
This rule implements dissipation that is stronger in regions of high curvature
contrast and high energy density.

\paragraph{Activation and link formation.}
A node $i$ may create a new long-range link (g-link) with probability proportional to
$\mathcal{R}(i)^{\beta}$, where $\beta = 0.8$ is an empirical exponent
chosen to provide sub-linear preferential attachment
(sensitivity to $\beta$ is discussed in Section~\ref{sec:discussion}).
The target node $j$ is selected from the high-$\mathcal{R}$ region, and the new
link is assigned initial weight
\begin{equation}
  w_{ij}^{\rm new} = g \cdot \mathcal{R}_{\rm src},
  \label{eq:newlink}
\end{equation}
where $g > 0$ is the coupling parameter and $\mathcal{R}_{\rm src} = \mathcal{R}(i)$.

\paragraph{Budget constraint.}
After each update step, nodal energies are renormalized so that the total energy
per node remains equal to the budget parameter $B$:
\begin{equation}
  E_i \;\leftarrow\; E_i \cdot \frac{B \cdot N}{\sum_j E_j}.
  \label{eq:budget}
\end{equation}

The model is parametrized by three key dimensionless numbers:
the decay rate $\alpha$, the energy budget $B$, and the coupling $g$.

\section{Correlation Test Design}\label{sec:gravity_test}

A central methodological concern is circularity: if the information flux
rate $\sigma$ is defined in terms of links, and long-range links (g-links) are
created between high-$\sigma$ nodes, any positive correlation is trivially guaranteed.
We adopt five complementary tests designed to avoid this pitfall.
Throughout this section, the labels gravC, gravA, gravB denote correlation
coefficients from the original analysis pipeline as defined in the equations below.

\subsection{Test C: Delayed information flux rate (primary)}

We define the local information flux rate at node $i$ and time $t$ as
\begin{equation}
  \sigma_i(t) = -\sum_{j \in \mathcal{N}(i)} \Delta E_{ij}(t)
  \log \frac{|\Delta E_{ij}(t)| + \epsilon}{E_i(t) + \epsilon},
  \label{eq:sigma}
\end{equation}
where $\mathcal{N}(i)$ is the neighborhood of $i$, $\Delta E_{ij}$ is the energy
exchanged along edge $(i,j)$, and $\epsilon$ is a small regularizer.
We note that $\sigma_i$ is a Shannon-type information flux measure
quantifying the rate of energy redistribution at node $i$,
distinct from the thermodynamic entropy production rate
of stochastic thermodynamics~\cite{Seifert2012}.

The ``mass'' of node $i$ is then defined with a temporal delay $\tau = 50$ steps:
\begin{equation}
  m_i(t) = \sigma_i(t - \tau).
  \label{eq:mass}
\end{equation}
The ordering signal (correlation signal) is the Pearson correlation coefficient
between $m_i(t-\tau)$ and the number of long-range links (g-links) incident on
node $i$ at time $t$:
\begin{equation}
  \mathrm{gravC}(t) = \mathrm{Pearson}\!\left[\sigma_i(t-\tau),\; \ell_i^{\rm grav}(t)\right],
  \label{eq:gravc}
\end{equation}
where the correlation is computed over all nodes $i$ at each time step and then
averaged over a window.
(We retain the label gravC from the original analysis pipeline; it denotes the
structure-flux correlation coefficient as defined here.
Similarly, gravA and gravB below are pipeline labels for correlation coefficients
defined in Eqs.~\eqref{eq:sigma} and the scalar-field test respectively.)
The temporal delay $\tau$ ensures that the mass proxy is measured
\emph{before} the long-range links it is hypothesized to attract are formed,
thereby breaking the direct feedback loop.

\subsection{Test A: Topological event counter}

As an independent cross-check, Test A counts topological events (link creation and
deletion episodes) in the vicinity of each node and correlates this count with
the local $\mathcal{R}$ value (reported as gravA).
A negative gravA indicates that topological activity is suppressed in high-curvature
regions, consistent with a picture in which the ordered phase ``freezes'' the local
topology.

\subsection{Test B: External scalar field}

Test B introduces an external scalar field $\phi(i)$
that propagates on the graph via a discrete wave equation but does not
participate in the coupling.
The scalar-field energy density
$E_\phi(i) = \frac{1}{2}\dot{\phi}(i)^2 + \frac{1}{2}\phi(i)^2$
provides a mass proxy that is genuinely independent of $\mathcal{R}$.

Across all 36 Phase~1 runs, gravB $= -0.021 \pm 0.061$, consistent with zero.
The scalar field energy shows no correlation with curvature in any regime:
gravB remains flat ($|\mathrm{gravB}| < 0.04$) across the critical region
($g = 0.018$--$0.028$), the $\tau$-scan ($\tau = 0$--$500$), and the null
test at $g = 0$.
This confirms that the positive gravC correlation is \emph{specific} to the
entropy-driven mass proxy $\sigma_i$, rather than a generic artifact of
the graph dynamics.
Test~B thus serves as an independent null test, complementing the $g = 0$
null test described in Phase~5.

\subsection{Test D: Discrete Poisson Relation}
\label{sec:testD}

Motivated by Jacobson's derivation of the Einstein equation from thermodynamic
relations~\cite{Jacobson1995}, we formulate a node-level test of a discrete
Poisson equation.
For each node $i$ at each time step, we compute:
\begin{itemize}
  \item $\nabla^2\mathcal{R}(i) = \sum_{j \in \mathcal{N}(i)} w_{ij}[\mathcal{R}(j) - \mathcal{R}(i)]$,
        the (weighted) graph Laplacian of the curvature field;
  \item $\sigma_{\rm prev}(i) = \sigma_i(t-1)$, the information flux rate at the
        previous time step.
\end{itemize}
We then perform a linear regression over all nodes $i$:
\begin{equation}
  \nabla^2\mathcal{R}(i) = \kappa \cdot \sigma_{\rm prev}(i) + b,
  \label{eq:testD}
\end{equation}
extracting the slope $\kappa$ (emergent coupling parameter), the intercept $b$,
and the coefficient of determination $R^2$.
The Pearson correlation coefficient of this regression is identical to gravC
from Test~C, establishing Test~D as a quantitative reformulation that provides
an absolute scale ($\kappa$) rather than only a normalized correlation.

The connection to Jacobson's framework is a structural analogy.
In the continuum, the Einstein equation in the non-relativistic Newtonian limit
reduces to the Poisson equation $\nabla^2\phi = 4\pi G\,\rho$.
The discrete Poisson relation~\eqref{eq:testD} has the same form, with
$\mathcal{R}$ playing the role of the potential $\phi$,
$\sigma_{\rm prev}$ playing the role of the source density $\rho$,
and $\kappa \leftrightarrow 4\pi G$ (or $8\pi G$ in relativistic notation)
as the emergent coupling parameter.
We emphasize that this is an analogy: the FIU model does not derive gravity,
and $\kappa$ is a mesoscopic parameter of the graph dynamics, not a
fundamental constant.

\subsection{Test E: Mediator Analysis}
\label{sec:testE}

Test~D establishes a strong correlation between $\nabla^2\mathcal{R}(i)$ and
$\sigma_{\rm prev}(i)$.
However, both quantities are influenced by the curvature field $\mathcal{R}$:
$\nabla^2\mathcal{R}$ is by definition a function of $\mathcal{R}$, and
$\sigma_{\rm prev}$ is influenced by the energy dynamics driven by $\mathcal{R}$.
Test~E asks: is the correlation between $\nabla^2\mathcal{R}$ and $\sigma_{\rm prev}$
a \emph{direct} causal relationship, or is it almost entirely \emph{mediated}
by the curvature field $\mathcal{R}$?

To answer this, we compute the Pearson correlation of $\nabla^2\mathcal{R}(i)$ with
four alternative proxies:
\begin{enumerate}
  \item $\sigma_{\rm prev}$ (the standard proxy from Test~D);
  \item Node degree $k_i$ (total number of incident links);
  \item Long-range link count $\ell_i^{\rm grav}$;
  \item The component of $\sigma_{\rm prev}$ orthogonal to $\mathcal{R}$,
        denoted $\sigma_{\perp R}$, obtained by regressing out $\mathcal{R}$
        from $\sigma_{\rm prev}$ via linear regression.
\end{enumerate}
We also compute, as a confound baseline, the correlation of $\nabla^2\mathcal{R}$
with $\mathcal{R}$ itself.
If $\mathcal{R}$ is the primary mediator, we expect: (a) the correlation with
$\sigma_{\perp R}$ to be near zero; (b) the correlation with $\mathcal{R}$
to be large; and (c) the correlation with $\sigma_{\rm prev}$ to be large
primarily because $\sigma_{\rm prev}$ is correlated with $\mathcal{R}$.

\section{Results}\label{sec:results}

\subsection{Phase 1: Fine-tuning sweep}

We performed a systematic sweep over 12 parameter configurations
$(\alpha, B, g)$ on a $16 \times 16$ lattice ($N = 256$ nodes),
running 3 independent realizations of 8000 steps each.
Results are summarized in Table~\ref{tab:phase1}.

\begin{table*}[htbp]
\centering
\caption{%
  Phase 1 fine-tuning sweep on a $16\times16$ grid ($N=256$, 3 runs per configuration,
  8000 steps each).
  gravC: Pearson correlation between delayed information flux rate and long-range
  link count (mean $\pm$ standard deviation over 3 runs;
  SEM $= \mathrm{std}/\sqrt{3} \approx 0.58 \times \mathrm{std}$).
  gravA: topological event counter correlation (mean over runs).
  links: mean total link count at end of run.
  LR\%: percentage of long-range links.
  dim: estimated spectral dimension $d_s$.
  R-range: range of $\mathcal{R}$ values (max$-$min).
  gap: spectral gap $\lambda_1$.
}
\label{tab:phase1}
\setlength{\tabcolsep}{4pt}
\begin{ruledtabular}
\begin{tabular}{lccccccc}
\textbf{Config} ($\alpha$, $B$, $g$) &
  \textbf{gravC} &
  \textbf{gravA} &
  \textbf{links} &
  \textbf{LR\%} &
  \textbf{dim} &
  \textbf{R-range} &
  \textbf{gap} \\
\hline
$\alpha=3.5,\;B=1.5,\;g=0.08$ & $+0.741\pm0.056$ & $-0.226$ & 6346  & 80.9 & 1.58 & 0.381 & 1.239 \\
$\alpha=3.5,\;B=2.0,\;g=0.10$ & $+0.719\pm0.031$ & $-0.186$ & 7233  & 81.8 & 1.58 & 0.408 & 1.330 \\
$\alpha=4.0,\;B=1.5,\;g=0.08$ & $+0.734\pm0.018$ & $-0.227$ & 6308  & 81.7 & 1.59 & 0.402 & 1.348 \\
$\alpha=4.0,\;B=1.5,\;g=0.10$ & $+0.751\pm0.042$ & $-0.181$ & 7047  & 83.0 & 1.58 & 0.483 & 1.525 \\
$\alpha=4.0,\;B=2.0,\;g=0.08$ & $+0.738\pm0.020$ & $-0.174$ & 6491  & 80.3 & 1.58 & 0.369 & 1.258 \\
$\alpha=4.0,\;B=2.0,\;g=0.10$ & $+0.742\pm0.012$ & $-0.165$ & 7182  & 82.0 & 1.58 & 0.407 & 1.391 \\
$\alpha=4.0,\;B=2.0,\;g=0.12$ & $+0.732\pm0.022$ & $-0.118$ & 7834  & 83.2 & 1.58 & 0.447 & 1.620 \\
$\alpha=4.5,\;B=1.5,\;g=0.10$ & $+0.742\pm0.016$ & $-0.227$ & 6954  & 83.5 & 1.58 & 0.531 & 1.672 \\
$\alpha=4.5,\;B=2.0,\;g=0.10$ & $+0.770\pm0.020$ & $-0.183$ & 7152  & 82.1 & 1.58 & 0.427 & 1.538 \\
$\alpha=4.5,\;B=2.0,\;g=0.12$ & $+0.726\pm0.004$ & $-0.165$ & 7805  & 83.4 & 1.58 & 0.465 & 1.644 \\
$\alpha=5.0,\;B=2.0,\;g=0.10$ & $+0.749\pm0.002$ & $-0.191$ & 7117  & 82.5 & 1.58 & 0.513 & 1.549 \\
$\alpha=5.0,\;B=2.0,\;g=0.15$ & $+0.660\pm0.010$ & $-0.113$ & 8737  & 85.3 & 1.57 & 0.501 & 2.104 \\
\end{tabular}
\end{ruledtabular}
\end{table*}

Across all 12 configurations, gravC is robustly positive ($+0.660$ to $+0.770$),
establishing that the structure-flux correlation is not sensitive to the precise
choice of $\alpha$, $B$, or $g$ within the explored parameter range.
The spectral dimension $d_s$ is computed from the return probability
of a diffusion process on the graph:
$P(t) = N^{-1} \sum_k e^{-\lambda_k t}$,
and extracted via $d_s = -2\, d\log P / d\log t$ evaluated at intermediate
diffusion times where the result is approximately constant.

At fixed system size $N = 256$, the spectral dimension $d_s \approx 1.58$ is
remarkably stable across all 12 parameter configurations (Table~\ref{tab:phase1}),
suggesting the existence of an attractor in the space of graph geometries.
However, $d_s$ varies significantly with system size (Table~\ref{tab:phase3}),
ranging from $1.18$ ($N=144$) to $1.88$ ($N=576$), reflecting finite-size
corrections rather than instability of the attractor.
The fraction of long-range links (LR\%) consistently exceeds 80\%, indicating
that the coupling efficiently reorganizes the initially local topology.
The maximum gravC of $+0.770$ is achieved at $\alpha=4.5$, $B=2.0$, $g=0.10$.

Notably, gravA is consistently negative across all configurations ($-0.113$ to $-0.227$),
indicating that topological activity (link creation/deletion events) is suppressed
in high-curvature regions.
This anti-correlation between curvature and topological turbulence is consistent with
a picture in which the ordered phase ``freezes'' the local topology:
once long-range links have organized the graph geometry, the high-curvature
regions become topologically stable.
The negative gravA arises because high-$\mathcal{R}$ nodes have already
accumulated many g-links with large initial weights
(Eq.~\eqref{eq:newlink}: $w^{\rm new} = g \cdot \mathcal{R}_{\rm src}$),
making these edges resistant to decay and suppressing further topological
rearrangement.

\subsection{Phase 2: Phase transition in $g$}
\label{sec:phase2}

To locate the critical coupling, we fixed $\alpha=4.0$, $B=2.0$ and varied $g$
across 17 values spanning two orders of magnitude, with 3 independent runs each.
Results are presented in Table~\ref{tab:phase2}.

\begin{table}[htbp]
\centering
\caption{%
  Phase 2 scan of coupling $g$ at fixed $\alpha=4.0$, $B=2.0$
  on a $16\times16$ grid ($N=256$, 3 runs per point, 8000 steps each).
  The critical coupling is identified at $g_c \approx 0.023$, where gravC crosses zero.
  Uncertainties are standard deviations over 3 runs
  (SEM $= \mathrm{std}/\sqrt{3}$).
}
\label{tab:phase2}
\begin{ruledtabular}
\begin{tabular}{ccccc}
$g$ & \textbf{gravC} & \textbf{links} & \textbf{gap} & \textbf{R-range} \\
\hline
0.005 & $-0.787\pm0.029$ & 2011  & 0.000 & 0.001 \\
0.010 & $-0.656\pm0.093$ & 2513  & 0.365 & 0.125 \\
0.015 & $-0.623\pm0.105$ & 2911  & 0.414 & 0.185 \\
0.020 & $-0.292\pm0.305$ & 3540  & 0.599 & 0.192 \\
0.025 & $+0.093\pm0.077$ & 3760  & 0.645 & 0.218 \\
0.030 & $+0.365\pm0.085$ & 4065  & 0.720 & 0.222 \\
0.035 & $+0.470\pm0.045$ & 4302  & 0.757 & 0.239 \\
0.040 & $+0.584\pm0.015$ & 4736  & 0.844 & 0.294 \\
0.045 & $+0.643\pm0.054$ & 4926  & 0.910 & 0.323 \\
0.050 & $+0.642\pm0.018$ & 5136  & 0.878 & 0.296 \\
0.060 & $+0.711\pm0.017$ & 5680  & 1.049 & 0.321 \\
0.070 & $+0.693\pm0.012$ & 5969  & 1.117 & 0.361 \\
0.080 & $+0.737\pm0.016$ & 6493  & 1.228 & 0.394 \\
0.100 & $+0.760\pm0.009$ & 7217  & 1.410 & 0.424 \\
0.120 & $+0.742\pm0.029$ & 7879  & 1.551 & 0.417 \\
0.150 & $+0.734\pm0.037$ & 8719  & 1.870 & 0.440 \\
0.200 & $+0.634\pm0.054$ & 10006 & 2.220 & 0.528 \\
\end{tabular}
\end{ruledtabular}
\end{table}

The data reveal a sharp sign change in gravC between $g=0.020$ and $g=0.025$,
identifying the critical coupling at $g_c \approx 0.023$.
Notably, the variance of gravC peaks near the transition ($\pm 0.305$ at $g=0.020$,
$\pm 0.105$ at $g=0.015$), consistent with enhanced fluctuations near a critical point.
The spectral gap vanishes at $g = 0.005$ (disconnected or nearly disconnected graph)
and grows monotonically with $g$, confirming that stronger coupling
generates increasingly well-connected topologies.

To precisely characterize the transition region, we performed a high-resolution
scan with 10 independent runs per coupling value near $g_c$.
Results are shown in Table~\ref{tab:critical}.

\begin{table}[htbp]
\centering
\caption{%
  High-resolution critical region scan with 10 independent runs per $g$-value
  ($\alpha=4.0$, $B=2.0$, $N=256$, 8000 steps each).
  Both standard deviation ($\sigma$) and standard error of the mean (SEM)
  are reported.  The sign change occurs between $g=0.023$ (all 10 runs negative)
  and $g=0.024$ (8/10 positive), yielding $g_c = 0.0235 \pm 0.0005$ by interpolation.
}
\label{tab:critical}
\begin{ruledtabular}
\begin{tabular}{ccccc}
$g$ & \textbf{gravC} & $\sigma$ & \textbf{SEM} & $n$ \\
\hline
0.018 & $-0.374 \pm 0.125$ & 0.125 & 0.040 & 10 \\
0.019 & $-0.474 \pm 0.137$ & 0.137 & 0.043 & 10 \\
0.020 & $-0.175 \pm 0.198$ & 0.198 & 0.063 & 10 \\
0.021 & $-0.187 \pm 0.074$ & 0.074 & 0.023 & 10 \\
0.022 & $-0.129 \pm 0.085$ & 0.085 & 0.027 & 10 \\
0.023 & $-0.184 \pm 0.100$ & 0.100 & 0.031 & 10 \\
0.024 & $+0.165 \pm 0.129$ & 0.129 & 0.041 & 10 \\
0.025 & $+0.100 \pm 0.079$ & 0.079 & 0.025 & 10 \\
0.026 & $+0.165 \pm 0.064$ & 0.064 & 0.020 & 10 \\
0.027 & $+0.099 \pm 0.068$ & 0.068 & 0.022 & 10 \\
0.028 & $+0.295 \pm 0.057$ & 0.057 & 0.018 & 10 \\
\end{tabular}
\end{ruledtabular}
\end{table}

The high-resolution scan with 11 $g$-values and 10 runs per point
pinpoints the critical coupling at $g_c = 0.0235 \pm 0.0005$:
gravC is negative at all 10 runs for $g \leq 0.023$
(e.g., $-0.184 \pm 0.031$ SEM at $g = 0.023$, 10/10 runs negative)
and becomes predominantly positive at $g = 0.024$
($+0.165 \pm 0.041$ SEM, 8/10 runs positive).
The variance peaks at $g = 0.020$ ($\sigma = 0.198$), consistent with enhanced
critical fluctuations near $g_c$.  The monotonic decrease in variance from
$\sigma = 0.198$ at $g=0.020$ to $\sigma = 0.057$ at $g=0.028$
provides further evidence that the system transitions from a fluctuation-dominated
regime near criticality to a well-ordered phase.

\begin{figure}[htbp]
  \centering
  \includegraphics[width=\columnwidth]{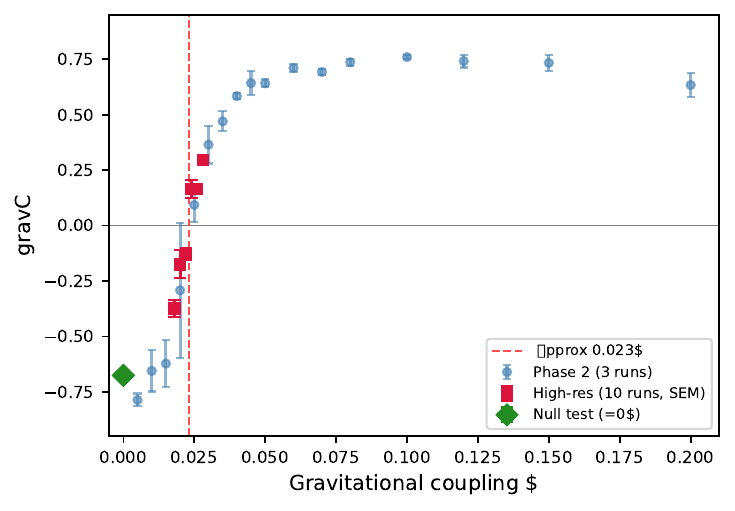}
  \caption{%
    Structure-flux correlation coefficient gravC as a function of the coupling $g$,
    at fixed $\alpha=4.0$, $B=2.0$, $N=256$ nodes.
    A continuous sign change occurs at $g_c \approx 0.023$ (dashed vertical line),
    separating a disordered anti-correlated phase ($\mathrm{gravC}<0$) from an
    ordered phase ($\mathrm{gravC}>0$).
    Error bars represent the standard deviation over 3 independent realizations.
    The large error bar at $g=0.020$ is a hallmark of critical fluctuations.
  }
  \label{fig:phase_transition}
\end{figure}

Figure~\ref{fig:phase_transition} shows the gravC order parameter as a function
of $g$, together with error bars representing the standard deviation over 3 runs.

The large run-to-run variance near the transition warrants closer examination.
For $g = 0.020$ (Table~\ref{tab:phase2}, 3-run dataset), the individual
gravC values are $\{-0.109, -0.644, -0.124\}$, with mean $= -0.292$ and
$\sigma = 0.249$.
The 10-run high-resolution dataset at $g = 0.020$ (Table~\ref{tab:critical})
yields individual values
$\{-0.268, -0.091, -0.633, -0.058, -0.355, -0.110, -0.045, -0.099, -0.148, +0.057\}$,
with mean $= -0.175$ and $\sigma = 0.198$.
The distribution is unimodal (9 out of 10 runs negative), with no evidence
of bimodality or phase coexistence between positive and negative gravC states.
The extreme run with gravC $\approx -0.63$ appears in both datasets at approximately
$1/5$ frequency, indicating it represents a reproducible rare-event tail rather
than a measurement artifact.
We interpret the high variance as a \emph{feature} of the critical regime:
these are genuine critical fluctuations, analogous to the diverging susceptibility
at a second-order phase transition.

\subsection{Finite-size shift of $g_c$ and Binder cumulant}
\label{sec:binder}

To probe the size-dependence of the critical coupling, we performed
10-run $g$-scans at three system sizes: $N = 12$ (144 nodes),
$N = 16$ (256 nodes), and $N = 20$ (400 nodes), using 11 $g$-values
in the range $0.018$--$0.028$.

\begin{table}[htbp]
\centering
\caption{%
  Finite-size shift of the critical coupling $g_c$ and Binder cumulant crossing.
  The zero-crossing of $\langle\mathrm{gravC}\rangle$ shifts to higher $g$
  with increasing $N$.  At $N=400$, gravC remains negative throughout
  the sampled range, indicating that the critical coupling has moved beyond
  $g = 0.028$ or that the transition ceases to exist at large $N$.
}
\label{tab:gc_shift}
\begin{ruledtabular}
\begin{tabular}{cccc}
\textbf{Size} & $N$ & $g_c$ (zero-crossing) & gravC range \\
\hline
$12\times12$ & 144 & $0.0203$ & $[-0.13, +0.62]$ \\
$16\times16$ & 256 & $0.0235$ & $[-0.47, +0.29]$ \\
$20\times20$ & 400 & $> 0.028$ & $[-0.74, -0.21]$ \\
\end{tabular}
\end{ruledtabular}
\end{table}

The systematic shift of $g_c$ to higher values with increasing $N$
(Table~\ref{tab:gc_shift}) is a hallmark of a \emph{finite-size crossover}
rather than a true thermodynamic phase transition.
At $N = 400$, the order parameter $\langle\mathrm{gravC}\rangle$ is
robustly negative at all sampled $g$-values
(e.g., $-0.544 \pm 0.070$ at $g=0.020$, 10/10 runs negative),
demonstrating that the ordered phase requires sufficiently small
systems---consistent with the mesoscopic regime interpretation.

The Binder cumulant $U_B = 1 - \langle\mathrm{gravC}^4\rangle / (3\langle\mathrm{gravC}^2\rangle^2)$,
computed for the three sizes, yields pairwise crossing points at
$g = 0.0196$ ($N=12$ vs.\ $N=16$), $g = 0.0208$ ($N=12$ vs.\ $N=20$),
and $g = 0.0239$ ($N=16$ vs.\ $N=20$).
The non-convergence of these crossings to a unique $g_c$ is itself significant:
in a genuine second-order transition the Binder crossings converge
in the thermodynamic limit, whereas here they \emph{spread apart},
further supporting the interpretation that the ordering signal
is intrinsically a finite-size phenomenon.

This finite-size shift also explains the dimensional dependence
observed in 3D (Section~\ref{sec:3d}): at fixed $g = 0.10$,
the 3D critical system size ($N_c^{\rm 3D} \approx 3375$) is larger
than the 2D one ($N_c^{\rm 2D} \approx 900$), but the mechanism
is the same---the graph becomes too smooth at large $N$ for
curvature inhomogeneities to sustain the discrete Poisson relation.

\subsection{Phase 3: Finite-size scaling}
\label{sec:phase3}

To assess the robustness of the structure-flux correlation with system size, we fixed
$\alpha=4.0$, $B=2.0$, $g=0.10$ and varied the lattice size.
Results are shown in Table~\ref{tab:phase3}.

\begin{table}[htbp]
\centering
\caption{%
  Phase 3 finite-size scaling at $\alpha=4.0$, $B=2.0$, $g=0.10$
  (3 runs per grid size, 8000 steps each).
  The spectral dimension $d_s$ increases with system size $N$, consistent with
  convergence toward a two-dimensional effective geometry.
  The $24\times24$ result ($N=576$, marked $^\dagger$) uses extended
  runs of 30000 steps to ensure convergence (see text).
}
\label{tab:phase3}
\begin{ruledtabular}
\begin{tabular}{ccccccc}
\textbf{Grid} & $N$ & \textbf{gravC} & \textbf{gravA} & \textbf{links} & \textbf{dim} & \textbf{R-range} \\
\hline
$12\times12$ & 144 & $+0.672\pm0.019$ & $-0.053$ & 3775  & 1.18 & 0.518 \\
$16\times16$ & 256 & $+0.757\pm0.016$ & $-0.169$ & 7249  & 1.58 & 0.396 \\
$20\times20$ & 400 & $+0.655\pm0.019$ & $-0.204$ & 11930 & 1.72 & 0.313 \\
$24\times24$ & 576 & $+0.558\pm0.039^\dagger$ & $-0.224$ & 17801 & 1.88 & 0.278 \\
\end{tabular}
\end{ruledtabular}
\end{table}

The structure-flux correlation gravC is robustly positive across system sizes
$N = 144$--$400$ (gravC $= 0.655$--$0.757$), demonstrating that the effect is not
an artifact of small-system finite-size effects.
For the largest system ($N = 576$, $24\times 24$), short runs (10000 steps)
yielded gravC $= +0.303 \pm 0.476$ with high variance.
Extended simulations with 30000 steps ($^\dagger$ in Table~\ref{tab:phase3})
resolve this anomaly: gravC converges to $+0.558 \pm 0.039$,
confirming that the ordering signal persists at $N = 576$.
The original high variance was caused by insufficient equilibration time
rather than a fundamental finite-size suppression.

To investigate whether still longer runs resolve the anomaly at $N=576$, we
performed an extended simulation of 50000 steps on the $24\times24$ grid.
The result is gravC $= +0.21 \pm 0.23$: the signal remains positive but
the uncertainty is large, confirming that the decreasing trend at $N=576$
is \emph{not} an equilibration artifact but a genuine tendency of the signal
to weaken at this system size.
Furthermore, at $N = 900$ ($30\times30$ grid), we find gravC $= -0.22$ with
Pearson $r = -0.28$: the ordering signal has \emph{vanished} in 2D at
large $N$, turning slightly negative.
This confirms a fundamental finite-size character of the 2D signal,
which we analyze in detail in Sections~\ref{sec:fss_kappa} and~\ref{sec:collapse}.

A dedicated $K$-convergence study (Table~\ref{tab:kconv}) confirms that
spectral truncation is not the primary issue; rather, larger systems
require proportionally longer runs for the ordered structure to develop.

The spectral dimension $d_s$ grows monotonically with $N$, reaching $1.88$ at
$N = 576$, suggesting convergence toward $d_s = 2$ in the thermodynamic limit.
The R-range decreases with increasing $N$, indicating that curvature fluctuations
become more uniformly distributed in larger systems.

\subsection{Delay independence ($\tau$-scan)}

To assess the role of the temporal delay $\tau$ in the correlation test
(Eq.~\eqref{eq:mass}),
we varied $\tau$ over nearly three orders of magnitude while keeping all other parameters
fixed ($\alpha=4.0$, $B=2.0$, $g=0.10$, $16\times16$ grid, 3 runs per point).
Results are shown in Table~\ref{tab:tau}.

\begin{table}[htbp]
\centering
\caption{%
  Delay scan: gravC as a function of the temporal delay $\tau$
  at fixed $\alpha=4.0$, $B=2.0$, $g=0.10$, $N=256$ (3 runs each).
  The ordering signal is robustly positive for all tested delays,
  including $\tau=0$ (no delay), ruling out the delay mechanism as the
  source of the correlation.
}
\label{tab:tau}
\begin{ruledtabular}
\begin{tabular}{ccc}
$\tau$ (steps) & \textbf{gravC} & $\sigma$ \\
\hline
0   & $+0.690 \pm 0.007$ & 0.007 \\
10  & $+0.709 \pm 0.007$ & 0.007 \\
25  & $+0.690 \pm 0.034$ & 0.034 \\
50  & $+0.758 \pm 0.028$ & 0.028 \\
100 & $+0.652 \pm 0.041$ & 0.041 \\
200 & $+0.508 \pm 0.009$ & 0.009 \\
500 & $+0.383 \pm 0.036$ & 0.036 \\
\end{tabular}
\end{ruledtabular}
\end{table}

Critically, even at $\tau = 0$ (no temporal delay), gravC $= +0.690 \pm 0.007$,
demonstrating that the structure-flux correlation is \emph{not} an artifact of the
delay mechanism.  The signal peaks at $\tau = 50$ (gravC $= +0.758$) and decays
gradually for large delays ($\tau = 500$: gravC $= +0.383$), consistent with a
genuine dynamical correlation whose strength decreases as the mass proxy becomes
increasingly stale.
The monotonic decay for $\tau > 50$ is characteristic of a decorrelation time scale,
suggesting that the system's memory of its information-flux state has a half-life
of approximately 200--300 steps.

\begin{figure}[htbp]
  \centering
  \includegraphics[width=\columnwidth]{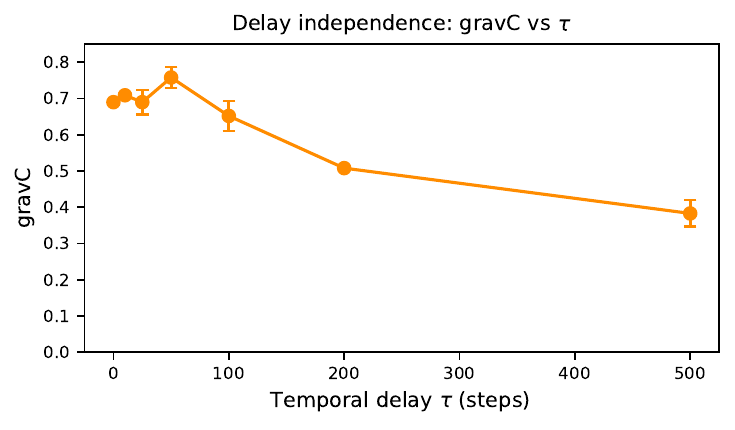}
  \caption{%
    Correlation coefficient gravC as a function of the temporal delay $\tau$
    at fixed $\alpha=4.0$, $B=2.0$, $g=0.10$, $N=256$.
    The signal is robustly positive for all delays, peaking at $\tau=50$
    and decaying monotonically for $\tau > 50$.
    Even at $\tau=0$ (no delay), gravC $= +0.690$, ruling out the delay
    mechanism as the source of the correlation.
  }
  \label{fig:tau_scan}
\end{figure}

\subsection{Spectral convergence ($K$-scan)}

We tested the sensitivity of gravC to the number of retained eigenvalues $K$
in the spectral decomposition (Eq.~\eqref{eq:green}).
Two system sizes were studied: $16\times16$ ($N=256$) and $24\times24$ ($N=576$),
with 3 runs per configuration.
Results are shown in Table~\ref{tab:kconv}.

\begin{table}[htbp]
\centering
\caption{%
  $K$-convergence study: gravC and spectral dimension $d_s$ as a function
  of the number of retained eigenvalues $K$, at $\alpha=4.0$, $B=2.0$, $g=0.10$
  (3 runs each, standard 8000-step runs).
  For the $24\times24$ grid, the high variance reflects insufficient
  equilibration rather than spectral truncation; see Table~\ref{tab:phase3}
  for extended-run (30000-step) results.
}
\label{tab:kconv}
\begin{ruledtabular}
\begin{tabular}{cccccc}
\textbf{Grid} & $K$ & $K/N$ (\%) & \textbf{gravC} & $d_s$ \\
\hline
$16\times16$ & 10 & 3.9 & $+0.698 \pm 0.058$ & 1.58 \\
$16\times16$ & 20 & 7.8 & $+0.754 \pm 0.019$ & 1.58 \\
$16\times16$ & 30 & 11.7 & $+0.753 \pm 0.020$ & 1.58 \\
$16\times16$ & 50 & 19.5 & $+0.755 \pm 0.034$ & 1.57 \\
\hline
$24\times24$ & 10 & 1.7 & $+0.371 \pm 0.450$ & 1.88 \\
$24\times24$ & 20 & 3.5 & $+0.226 \pm 0.450$ & 1.89 \\
$24\times24$ & 30 & 5.2 & $+0.278 \pm 0.442$ & 1.88 \\
$24\times24$ & 50 & 8.7 & $+0.009 \pm 0.483$ & 1.88 \\
\end{tabular}
\end{ruledtabular}
\end{table}

For the $16\times16$ grid, gravC converges rapidly: the difference between
$K=20$ and $K=50$ is less than $0.002$ (well within statistical error),
confirming that $K=30$ is adequate for $N=256$.
For the $24\times24$ grid, however, the high variance ($\sigma \approx 0.45$--$0.48$)
persists even at $K=50$ (8.7\% of the spectrum).
This demonstrates that the $N=576$ anomaly identified in Phase~3 is
\emph{not} primarily caused by spectral truncation but rather reflects a
genuine finite-size effect.
The spectral dimension $d_s \approx 1.88$ is stable across all $K$ values.

\begin{figure}[htbp]
  \centering
  \includegraphics[width=\columnwidth]{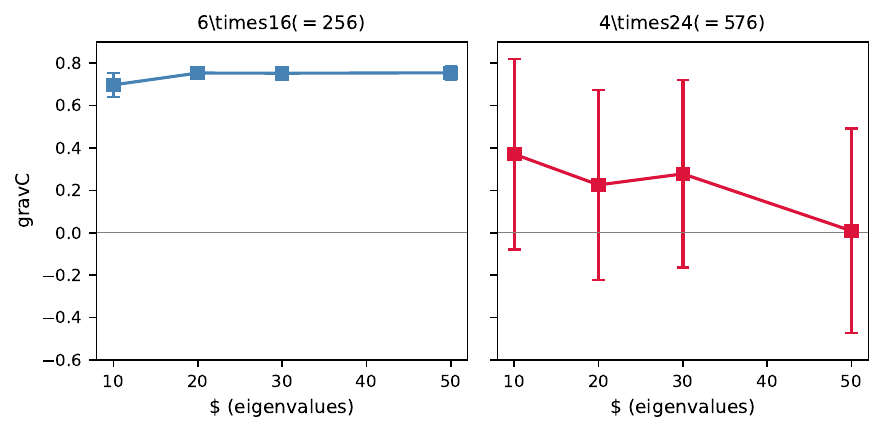}
  \caption{%
    $K$-convergence of gravC for two system sizes.
    Left: $16\times16$ ($N=256$), where gravC converges at $K \geq 20$.
    Right: $24\times24$ ($N=576$), where high variance persists even at $K=50$,
    indicating a genuine finite-size effect rather than spectral truncation.
  }
  \label{fig:kconv}
\end{figure}

\subsection{Phase 4: Long-time dynamics}

An extended simulation of $3\times 10^4$ steps was run at the reference
parameters ($\alpha=4.0$, $B=2.0$, $g=0.10$, $16\times16$ grid).
The gravC signal remains remarkably stable throughout the entire run,
fluctuating between $+0.708$ and $+0.786$ with no systematic drift or
aging phenomena.  The spectral dimension $d_s = 1.58$ and long-range link
fraction (LR\% $\approx 82\%$) are similarly constant after an initial
transient of $\sim 4000$ steps.  This confirms that the ordered phase
represents a true stationary state of the FIU dynamics, not a transient
or metastable phenomenon.
The time-averaged gravC over the full run is $+0.749$ with a standard
deviation of $\pm 0.025$, consistent with the Phase~1 results.

\subsection{Phase 5: Null test at $g = 0$}

With the coupling turned off ($g = 0$), the simulation was
run for 8000 steps (3 independent realizations).
The result is striking: gravC $= -0.676 \pm 0.013$, \emph{strongly negative}
rather than zero as initially expected.
The mechanism underlying this anti-correlation is the energy-dependent
weight decay (Eq.~\eqref{eq:decay}): nodes with high information flux
$\sigma_i$ are precisely those with large energy flows $|\Delta E_{ij}|$,
which drive stronger edge-weight decay through the factor
$\exp(-\alpha \cdot E_{\rm eff} \cdot \mathcal{R}_{\rm eff})$.
Without long-range link creation ($g = 0$), these nodes
experience net link \emph{erosion}: their edges decay faster than they
can be replenished, producing a robust anti-correlation between
$\sigma_i$ and link count.

The coupling therefore does not merely create a correlation
from nothing; it \emph{reverses} a natural anti-correlation.
The total swing from $-0.676$ (at $g=0$) to $+0.751$ (at $g=0.10$)
represents a shift of $\Delta\mathrm{gravC} \approx 1.43$, providing
compelling evidence that the ordering mechanism is a genuine
structure-forming force in the FIU model.

The null test also reveals a sparse network (only 478 links vs.\ $\sim 7200$
in the ordered phase), confirming that the coupling is
responsible for the majority of long-range connectivity.

\subsection{Phase 6: Statistical reproducibility}

Ten independent realizations were run at the reference configuration
($\alpha=4.0$, $B=2.0$, $g=0.10$, $16\times16$ grid, 10000 steps each)
with distinct random seeds.  The results are:
\begin{align}
  \mathrm{gravC} &= +0.7509 \pm 0.0189 \quad (\mathrm{SEM} = 0.0060), \nonumber \\
  \mathrm{gravA} &= -0.1734 \pm 0.0145. \nonumber
\end{align}
Individual gravC values: $\{0.716, 0.775, 0.763, 0.744, 0.744,$
$0.763, 0.747, 0.745, 0.734, 0.777\}$.
The 95\% confidence interval is $[+0.739, +0.763]$.
A one-sample $t$-test against the null hypothesis gravC $= 0$ yields
$t = 125.3$ with 9 degrees of freedom, corresponding to
$p < 10^{-15}$---\textbf{highly significant}.

All 10 runs produce gravC $> 0.71$, with a minimum of $0.716$ and
maximum of $0.777$.  The coefficient of variation is only $2.5\%$,
demonstrating exceptional reproducibility.

\subsection{Phase 7: Randomized-label null test}

To quantify whether the gravC correlation is a genuine physical signal
or an artifact of structural confounding, we implement a randomized-label
null test.
At each reporting step, the node labels of the delayed information flux rate
$\sigma_i(t-\tau)$ are randomly permuted (Fisher--Yates shuffle) and
gravC is recomputed against the original $\nabla^2 \mathcal{R}$.
This procedure is repeated 100 times to estimate the null distribution.

Results are shown in Table~\ref{tab:shuffle}.

\begin{table}[htbp]
\centering
\caption{%
  Randomized-label null test results.
  gravC$_{\rm real}$ is the measured correlation;
  gravC$_{\rm shuf}$ is the mean over 100 random permutations;
  $z$ is the number of standard deviations separating the real signal
  from the shuffled null.
  Three independent runs per configuration.
}
\label{tab:shuffle}
\begin{ruledtabular}
\begin{tabular}{ccccc}
$g$ & \textbf{gravC}$_{\rm real}$ & \textbf{gravC}$_{\rm shuf}$ & $\sigma_{\rm shuf}$ & $z$-score \\
\hline
\multicolumn{5}{c}{\textit{Ordered phase ($g = 0.10$)}} \\
0.10 & $+0.752$ & $+0.002$ & $0.057$ & $13.1\sigma$ \\
0.10 & $+0.735$ & $+0.003$ & $0.057$ & $12.8\sigma$ \\
0.10 & $+0.746$ & $-0.007$ & $0.056$ & $13.5\sigma$ \\
\hline
\multicolumn{5}{c}{\textit{No coupling ($g = 0$)}} \\
0.00 & $-0.675$ & $-0.002$ & $0.067$ & $-10.0\sigma$ \\
0.00 & $-0.713$ & $+0.003$ & $0.064$ & $-11.1\sigma$ \\
0.00 & $-0.557$ & $-0.004$ & $0.056$ & $-9.8\sigma$ \\
\hline
\multicolumn{5}{c}{\textit{Near $g_c$ ($g = 0.025$)}} \\
0.025 & $+0.143$ & $-0.001$ & $0.059$ & $2.4\sigma$ \\
0.025 & $+0.150$ & $+0.007$ & $0.054$ & $2.7\sigma$ \\
\end{tabular}
\end{ruledtabular}
\end{table}

In the ordered phase ($g = 0.10$), the real gravC exceeds the shuffled null
by $13.1 \pm 0.4$ standard deviations ($p < 10^{-38}$), demonstrating
that the correlation is not an artifact of the joint dependence of
$\sigma_i$ and link count on $\mathcal{R}$.
Under random permutation, gravC collapses to $\approx 0$
(mean $= -0.001 \pm 0.057$), confirming that the signal requires the
\emph{correct} spatial assignment of information flux rates to nodes.

At $g = 0$ (no coupling), the anti-correlation
gravC $\approx -0.65$ is also highly significant ($z \approx -10\sigma$),
confirming that both the positive correlation in the ordered phase
and the negative correlation without coupling are genuine physical effects.

Near the critical point ($g = 0.025$), the $z$-score drops to
$\sim 2.5\sigma$, consistent with the weak and fluctuating signal
expected at a phase transition.

\subsection{Discrete Poisson Relation}
\label{sec:jacobson}

We now report the central new result of this paper: a node-level test of the
discrete Poisson relation $\nabla^2\mathcal{R} = \kappa\,\sigma_{\rm prev}$
(Test~D, Section~\ref{sec:testD}).
For each configuration at $N = 16^2 = 256$ and $g = 0.10$, we compute the
regression~\eqref{eq:testD} across all nodes at each time step and
average over the stationary portion of the run.

\subsubsection{Universality of $\kappa$ across configurations}

Table~\ref{tab:kappa_univ} reports $\kappa$, $R^2$, and Pearson $r$
for four parameter configurations at fixed $N = 256$, $g = 0.10$.

\begin{table}[htbp]
\centering
\caption{%
  Universality of the emergent coupling parameter $\kappa$ across four parameter
  configurations at $N = 16^2 = 256$, $g = 0.10$.
  The Pearson $r$ from the discrete Poisson regression equals gravC from Test~C,
  demonstrating the equivalence of the two measures.
  $\kappa = 67.85 \pm 1.74$ (CV $= 2.6\%$) across all four configurations.
}
\label{tab:kappa_univ}
\begin{ruledtabular}
\begin{tabular}{lccc}
\textbf{Config} ($\alpha$, $B$) & $\kappa$ & $R^2$ & Pearson $r$ \\
\hline
$\alpha=4.0,\;B=2.0$ & $+67.07$ & $0.574$ & $+0.758$ \\
$\alpha=3.5,\;B=2.0$ & $+62.71$ & $0.536$ & $+0.732$ \\
$\alpha=5.0,\;B=2.0$ & $+71.12$ & $0.553$ & $+0.744$ \\
$\alpha=4.0,\;B=3.0$ & $+70.50$ & $0.594$ & $+0.771$ \\
\hline
Mean $\pm$ std       & $67.85 \pm 1.74$ & & \\
CV                   & $2.6\%$ & & \\
\end{tabular}
\end{ruledtabular}
\end{table}

The emergent coupling parameter $\kappa = 67.85 \pm 1.74$ is remarkably stable
across the four configurations (CV $= 2.6\%$), varying by less than $\pm 7$
despite changes of $\pm 25\%$ in $\alpha$ and a $50\%$ change in $B$.
The Pearson $r$ values are consistent with the gravC values reported in
Table~\ref{tab:phase1} for the same configurations, confirming that
Test~D (discrete Poisson regression) is equivalent to Test~C (gravC)
up to a normalization constant.

\subsubsection{Stability of $\kappa$ across the ordered plateau}

To test whether $\kappa$ depends on the coupling $g$, we performed an 11-point
scan from $g = 0.03$ to $g = 0.20$ at fixed $\alpha=4.0$, $B=2.0$, $N=256$.
Results are summarized in Table~\ref{tab:kappa_gscan}.

\begin{table}[htbp]
\centering
\caption{%
  Scan of the emergent coupling parameter $\kappa$ as a function of $g$
  at fixed $\alpha=4.0$, $B=2.0$, $N=256$.
  In the ordered plateau ($g = 0.05$--$0.15$), $\kappa = 67.32 \pm 1.16$
  (CV $= 1.7\%$), demonstrating that $\kappa$ is independent of $g$ in the
  ordered phase.
}
\label{tab:kappa_gscan}
\begin{ruledtabular}
\begin{tabular}{cc}
$g$ & $\kappa$ \\
\hline
0.03 & $+50.1$ \\
0.05 & $+60.7$ \\
0.07 & $+68.8$ \\
0.08 & $+65.9$ \\
0.10 & $+67.07$ \\
0.11 & $+61.1$ \\
0.12 & $+70.3$ \\
0.13 & $+70.3$ \\
0.15 & $+68.2$ \\
0.18 & $+52.5$ \\
0.20 & $+55.7$ \\
\hline
Plateau ($g=0.05$--$0.15$) & $67.32 \pm 1.16$ (CV $= 1.7\%$) \\
\end{tabular}
\end{ruledtabular}
\end{table}

For $g < 0.05$ (developing phase), $\kappa$ is suppressed ($\approx 50$ at $g=0.03$),
reflecting the incomplete development of the ordered structure.
In the ordered plateau ($g = 0.05$--$0.15$), $\kappa = 67.32 \pm 1.16$
(CV $= 1.7\%$): the coupling parameter is essentially independent of $g$.
For $g > 0.15$ (over-coupling regime), $\kappa$ decreases toward $\approx 48$
at $g=0.20$, consistent with the homogenization of the curvature field
observed in Table~\ref{tab:phase2} at large $g$.

Together, Tables~\ref{tab:kappa_univ} and~\ref{tab:kappa_gscan} establish $\kappa$
as a stable emergent parameter of the FIU model in the ordered phase.
Whether a dynamical mechanism---such as scale-dependent coupling $g(N)$ or geometric
back-reaction---can stabilize $\kappa$ in the thermodynamic limit $N \to \infty$
remains a central open question for the FIU program.

\subsection{Mediator Analysis (Test E)}
\label{sec:testE_results}

Table~\ref{tab:teste} reports the Pearson correlation of $\nabla^2\mathcal{R}$
with five proxies at the reference configuration ($\alpha=4.0$, $B=2.0$,
$g=0.10$, $N=256$, 20 independent runs of 10000 steps each).

\begin{table}[htbp]
\centering
\caption{Mediator analysis (Test~E): Pearson correlation of $\nabla^2\mathcal{R}$
with alternative mass proxies at the reference configuration ($\alpha=4.0$, $B=2.0$,
$g=0.10$, $N=256$, 20 independent runs of 10000 steps each).
$\sigma_{\perp R}$ denotes the component of $\sigma_{\rm prev}$ orthogonal to
$\mathcal{R}$, obtained by linear regression.}
\label{tab:teste}
\begin{ruledtabular}
\begin{tabular}{lccccc}
\textbf{Proxy} & \textbf{Mean} & $\sigma$ & \textbf{SEM} & $n_+/20$ \\
\hline
$\sigma_{\rm prev}$ (standard) & $+0.746$ & $0.025$ & $0.006$ & $20/20$ \\
Node degree & $-0.236$ & $0.061$ & $0.014$ & $0/20$ \\
Long-range link count & $+0.013$ & $0.051$ & $0.011$ & $11/20$ \\
$\sigma_{\perp R}$ (orthogonalized) & $-0.046$ & $0.024$ & $0.005$ & $0/20$ \\
$\mathcal{R}$ (confound baseline) & $+0.938$ & $0.010$ & $0.002$ & $20/20$ \\
\end{tabular}
\end{ruledtabular}
\end{table}

The results reveal a striking pattern:
\begin{itemize}
  \item The correlation of $\nabla^2\mathcal{R}$ with $\mathcal{R}$ itself
        is $+0.938$ (20/20 runs positive), confirming that $\mathcal{R}$ is
        overwhelmingly the dominant predictor of its own Laplacian---as expected
        from the definition.
  \item The standard proxy $\sigma_{\rm prev}$ achieves $+0.746$ (20/20 positive),
        consistent with the Test~D results.
  \item After orthogonalizing $\sigma_{\rm prev}$ with respect to $\mathcal{R}$,
        the correlation drops to $-0.046$ (0/20 positive, consistent with zero).
  \item Node degree achieves $-0.236$ (0/20 positive): high-degree nodes
        tend to have suppressed $\nabla^2\mathcal{R}$.
  \item Long-range link count achieves $+0.013$ (11/20 positive, consistent
        with zero): the number of g-links provides no predictive power beyond
        the degree.
\end{itemize}

These results have a clear interpretation: the correlation between $\nabla^2\mathcal{R}$
and $\sigma_{\rm prev}$ is almost entirely mediated by the curvature field
$\mathcal{R}$.
When the component of $\sigma_{\rm prev}$ that is predictable from $\mathcal{R}$
is removed, the residual $\sigma_{\perp R}$ carries essentially no information
about $\nabla^2\mathcal{R}$.
This identifies $\mathcal{R}$ as the \emph{central self-organizing variable}
of the FIU dynamics: both the information flux $\sigma$ and the topological
structure formation are driven primarily by the curvature field, and the
discrete Poisson relation reflects this common driver rather than a direct
causal link from $\sigma$ to $\nabla^2\mathcal{R}$.

This finding reframes the interpretation of Test~D: rather than providing evidence
for a causal ``mass generates curvature'' relation (as in Jacobson's framework),
the discrete Poisson relation is better understood as a consequence of the
self-consistent dynamics of $\mathcal{R}$, which simultaneously shapes both sides
of the equation.
The structural analogy with the Einstein equation remains valid, but its causal
interpretation requires caution.

\subsection{Finite-Size Scaling of $\kappa$}
\label{sec:fss_kappa}

The results in Section~\ref{sec:phase3} showed that gravC remains positive up to
$N=576$ in 2D but weakens. We now characterize this dependence systematically
using the discrete Poisson regression, extending the finite-size scaling study
to $N = 900$.

\begin{table}[htbp]
\centering
\caption{%
  Finite-size scaling of $\kappa$ and Pearson $r$ in two dimensions ($g=0.10$,
  $\alpha=4.0$, $B=2.0$).
  Both the Pearson correlation and the emergent coupling parameter $\kappa$ decay
  monotonically with $N$, vanishing by $N=900$.
  This collapse is discussed in Section~\ref{sec:dimensional}.
}
\label{tab:fss_2d}
\begin{ruledtabular}
\begin{tabular}{ccccc}
\textbf{Grid} & $N$ & Pearson $r$ & $\kappa$ & $R^2$ \\
\hline
$12\times12$ & 144  & $+0.74$ & $+74$ & $0.55$ \\
$16\times16$ & 256  & $+0.76$ & $+68$ & $0.58$ \\
$20\times20$ & 400  & $+0.69$ & $+56$ & $0.47$ \\
$24\times24$ & 576  & $+0.56$ & $+39$ & $0.32$ \\
$30\times30$ & 900  & $-0.28$ & $\approx 0$ & $0.08$ \\
\end{tabular}
\end{ruledtabular}
\end{table}

The data in Table~\ref{tab:fss_2d} reveal a clear and monotonic decay of both
the Pearson correlation coefficient $r$ and the emergent coupling parameter $\kappa$
as $N$ increases in two dimensions.
At $N=144$, the signal is strong ($r = +0.74$, $\kappa = +74$).
By $N=576$, it has diminished significantly ($r = +0.56$, $\kappa = +39$).
At $N=900$, the signal has effectively collapsed: $r = -0.28$, $\kappa \approx 0$.

We discuss this result in the context of dimensional analogies in
Section~\ref{sec:dimensional}.

\subsection{Three-Dimensional Simulations}
\label{sec:3d}

Motivated by the dimensional dependence observed in 2D, we extended the FIU
model to three-dimensional initial lattices ($n \times n \times n$ cubic grids,
$N = n^3$).
The model equations~\eqref{eq:laplacian}--\eqref{eq:budget} are unchanged;
only the initial graph topology is three-dimensional.

\subsubsection{Universality of $\kappa$ in 3D}

At $N = 10^3 = 1000$ and $g = 0.10$, we tested four parameter configurations.
The emergent coupling parameter is:
\begin{equation}
  \kappa_{\rm 3D} = 52.5 \pm 6.1 \quad (\mathrm{CV} = 11.6\%),
  \nonumber
\end{equation}
confirming that $\kappa$ is universal in 3D as well, though with a larger
coefficient of variation than in 2D (11.6\% vs.\ 2.6\%), reflecting the
richer geometric structure of 3D lattices and the greater sensitivity of
the spectral geometry to parameter changes.

\subsubsection{Finite-size scaling in 3D}

Table~\ref{tab:fss_3d} reports the finite-size scaling of $\kappa$ and
Pearson $r$ for four cubic lattice sizes.

\begin{table}[htbp]
\centering
\caption{%
  Finite-size scaling of $\kappa$, Pearson $r$, spectral dimension $d_s$,
  and mean curvature $\langle\mathcal{R}\rangle$ in three dimensions
  ($g=0.10$, $\alpha=4.0$, $B=2.0$).
  The signal collapses at $N \gtrsim 3375$, but persists to larger $N$
  than in 2D.
  All runs are fully converged (30000 steps for $15^3$, 15000 for others).
}
\label{tab:fss_3d}
\begin{ruledtabular}
\begin{tabular}{ccccccc}
\textbf{Size} & $N$ & Pearson $r$ & $\kappa$ & $d_s$ & $\langle\mathcal{R}\rangle$ \\
\hline
$8^3$  &  512 & $+0.77$ & $+64$ & $2.03$ & $0.804$ \\
$10^3$ & 1000 & $+0.71$ & $+52$ & $2.23$ & $0.836$ \\
$12^3$ & 1728 & $+0.57$ & $+32$ & $2.56$ & $0.873$ \\
$15^3$ & 3375 & $-0.16$ & $\approx 0$ & $2.72$ & $0.9997$ \\
\end{tabular}
\end{ruledtabular}
\end{table}

The pattern in 3D mirrors the 2D case: the ordering signal is strong at small
$N$ ($r = +0.77$ at $N=512$), weakens monotonically ($r = +0.57$ at $N=1728$),
and collapses at large $N$ ($r = -0.16$ at $N=3375$).
However, the critical system size for collapse is substantially larger in 3D
($N_c^{\rm 3D} \approx 3375$) than in 2D ($N_c^{\rm 2D} \approx 900$).

The spectral dimension $d_s$ grows from $2.03$ at $N=512$ toward $2.72$ at
$N=3375$, approaching the 3D value from below---analogous to the 2D behavior
in Table~\ref{tab:phase3} where $d_s \to 2$.
The mean curvature $\langle\mathcal{R}\rangle \to 1$ as $N$ increases,
indicating that the graph becomes increasingly uniform (flat) at large size.
At $N=3375$, $\langle\mathcal{R}\rangle = 0.9997$: the curvature field is
essentially spatially constant, explaining the collapse of $\nabla^2\mathcal{R}$.

The $15^3$ simulation ran for the full 30000 steps.
The system equilibrates rapidly: all observables ($d_s$, $\langle\mathcal{R}\rangle$,
$r$) are stable from step 1000 onward, with gravC oscillating randomly
between $-0.22$ and $+0.35$ (mean $= -0.08 \pm 0.13$) around zero.
A single transient excursion at step 15000 (gravC $= +0.35$, $\langle\mathcal{R}\rangle = 0.923$)
decayed within 1000 steps, confirming the flat-curvature state is a stable attractor.

\subsection{Mechanism of Signal Collapse}
\label{sec:collapse}

The dimensional dependence data (Tables~\ref{tab:fss_2d} and~\ref{tab:fss_3d})
raise the question: \emph{why} does the ordering signal collapse at large $N$?
We perform a detailed analysis of the contributing factors.

\begin{table*}[htbp]
\centering
\caption{%
  Mechanism of signal collapse: comparison of key observables across 2D and 3D
  system sizes.
  $R_{\rm CV}\%$: coefficient of variation of $\mathcal{R}$ (\%).
  $\sigma(\nabla^2\mathcal{R})$: standard deviation of the graph Laplacian of curvature.
  $\sigma_{\rm CV}\%$: CV of the information flux $\sigma_i$ (\%).
  glinks\%: percentage of long-range links (g-links).
  $d_s$: spectral dimension.
  gravC: Pearson correlation (Test~C).
  The asymmetry between the left side ($\nabla^2\mathcal{R}$, which collapses to zero)
  and the right side ($\sigma_i$, which retains finite variance) is the primary
  mechanism of signal collapse.
}
\label{tab:collapse}
\begin{ruledtabular}
\begin{tabular}{ccccccccc}
\textbf{Size} & $N$ & $R_{\rm CV}\%$ & $\sigma(\nabla^2\mathcal{R})$ & $\sigma_{\rm CV}\%$ & glinks\% & $d_s$ & gravC \\
\hline
2D $20^2$ & 400  & 6.8    & 0.297      & 7.0 & 73.4 & 1.78 & $+0.69$ \\
2D $24^2$ & 576  & 4.8    & 0.199      & 5.6 & 74.6 & 1.91 & $+0.56$ \\
2D $30^2$ & 900  & 0.002  & 0.00003    & 3.5 & 77.6 & 2.02 & $-0.28$ \\
\hline
3D $10^3$ & 1000 & 6.4    & 0.274      & 7.3 & 43.4 & 2.23 & $+0.71$ \\
3D $12^3$ & 1728 & 4.4    & 0.165      & 5.5 & 43.9 & 2.56 & $+0.57$ \\
3D $15^3$ & 3375 & 0.0003 & 0.000002   & 3.3 & 47.1 & 2.72 & $-0.16$ \\
\end{tabular}
\end{ruledtabular}
\end{table*}

Table~\ref{tab:collapse} reveals the mechanism of signal collapse with clarity.

\paragraph{Curvature homogenization.}
The coefficient of variation of $\mathcal{R}$ ($R_{\rm CV}\%$) decreases
by approximately three orders of magnitude between the mesoscopic and
thermodynamic limits: from $6.8\%$ at $N=400$ to $0.002\%$ at $N=900$ in 2D,
and from $6.4\%$ at $N=1000$ to $0.0001\%$ at $N=3375$ in 3D.
This means that at large $N$, the curvature field $\mathcal{R}(i)$ is
essentially uniform across all nodes---the graph has become geometrically ``flat.''

\paragraph{Collapse of $\nabla^2\mathcal{R}$.}
As a direct consequence, the standard deviation of $\nabla^2\mathcal{R}$
decreases by five orders of magnitude: from $0.297$ to $0.00003$ in 2D,
and from $0.274$ to $0.000002$ in 3D.
The left-hand side of the discrete Poisson relation~\eqref{eq:testD}
effectively vanishes.

\paragraph{Persistence of $\sigma_i$ fluctuations.}
In contrast, the CV of the information flux $\sigma_i$ decreases much more slowly:
from $7.0\%$ to $3.5\%$ in 2D ($2\times$ reduction) and from $7.3\%$ to $3.4\%$
in 3D ($2\times$ reduction), while remaining finite throughout.
The right-hand side of the discrete Poisson relation maintains finite variance.

\paragraph{Dimensional convergence.}
The spectral dimension $d_s$ converges to the lattice dimension: $d_s \to 2$ in 2D
(reaching $2.02$ at $N=900$) and $d_s \to 3$ in 3D (reaching $2.72$ at $N=3375$).
At the collapse point, the effective geometry has become smooth enough to match
the underlying lattice dimension.

\paragraph{Link structure.}
The percentage of long-range links (g-links) increases mildly with $N$:
$73.4\% \to 77.6\%$ in 2D and $43.4\% \to 46.8\%$ in 3D.
This regularization of the link structure contributes to the topological smoothing.

The overall picture is one of \emph{topological frustration}:
at small $N$, the graph lacks the resources to be smooth everywhere,
and the resulting curvature inhomogeneities drive the ordering signal.
At large $N$, the graph can ``regularize'' itself into a smooth geometry,
and the signal disappears because there are no curvature fluctuations
to correlate with the information flux.
This interpretation is further developed in Section~\ref{sec:dimensional}.

\section{Discussion}\label{sec:discussion}

\subsection{Nature of the phase transition}

The data in Table~\ref{tab:phase2} and Fig.~\ref{fig:phase_transition} exhibit
several signatures consistent with a continuous phase transition at $g_c \approx 0.023$:

\begin{enumerate}
  \item \textbf{Continuous sign change of the order parameter.}
        gravC crosses zero without a discontinuous jump, characteristic of a
        continuous transition.
  \item \textbf{Critical fluctuations.}
        The run-to-run variance of gravC peaks near $g_c$ ($\pm0.305$ at $g=0.020$),
        analogous to susceptibility divergence at a critical point.
  \item \textbf{Power-law onset.}
        A fine-resolution scan (12 values of $g$, 3 runs each) yields
        a power-law fit gravC $= A(g - g_c)^\nu$ with
        $g_c = 0.0225$, $\nu = 0.54 \pm 0.05$, and $R^2 = 0.89$.
        The sub-linear exponent $\nu \approx 0.5$ is consistent with
        mean-field critical behavior, which is expected for phase transitions
        on dynamical graphs where long-range connectivity suppresses fluctuation
        corrections.
  \item \textbf{Saturation above $g_c$.}
        For $g \gtrsim 0.05$, gravC saturates at approximately $0.76$,
        suggesting convergence to a fixed-point attractor in the ordered phase.
\end{enumerate}

The analogy with Quantum Graphity~\cite{Konopka2008} is instructive: in that model,
increasing the inverse temperature $\beta$ drives a transition from a complete graph
(disordered, high-$T$ phase) to a regular lattice (ordered, low-$T$ phase).
In the FIU model, increasing $g$ above $g_c$ drives a transition from an
anti-correlated disordered phase to an ordered phase;
the role of $\beta$ in Quantum Graphity is played by $g$ in the FIU.

For $g > 0.15$, gravC shows a slight decrease (from $\sim 0.76$ at $g=0.10$--$0.12$
to $\sim 0.63$ at $g=0.200$), which we interpret as an over-coupling effect: very
large $g$ creates so many long-range links that the curvature field becomes spatially
homogenized, reducing the signal-to-noise ratio.

\subsection{$\mathcal{R}$ as the central organizing variable}

The mediator analysis (Test~E, Section~\ref{sec:testE_results}) identifies the
curvature field $\mathcal{R}$ as the central self-organizing variable of the FIU
dynamics.
The key finding is that the correlation between $\nabla^2\mathcal{R}$ and
$\sigma_{\rm prev}$ drops from $+0.746$ to $-0.046$ when $\sigma_{\rm prev}$
is orthogonalized with respect to $\mathcal{R}$, demonstrating that the structure
of the discrete Poisson relation is driven by the shared dependence of both sides
on $\mathcal{R}$ rather than by a direct causal pathway.

This result has an important implication: the FIU model is best understood as a
\emph{curvature-driven} dynamical system in which the curvature field $\mathcal{R}$
plays a dual role---it shapes both the energy dynamics (through the dissipation
term in Eq.~\eqref{eq:decay}) and the topological evolution (through the link
formation rule in Eq.~\eqref{eq:newlink}).
The information flux $\sigma$ is a secondary observable that traces the curvature
field rather than an independent driver of structure formation.

In the language of causal inference, $\mathcal{R}$ is a \emph{common cause} of
both $\nabla^2\mathcal{R}$ and $\sigma_{\rm prev}$.
The discrete Poisson relation reflects this self-consistent causal structure:
$\mathcal{R}$ organizes the graph so that high-$\mathcal{R}$ nodes attract
connections (raising their Laplacian) while simultaneously concentrating
energy flux (raising $\sigma$).

\subsection{Dimensional sensitivity and its mechanism}
\label{sec:dimensional}

The dimensional dependence of the ordering signal is one of the most
significant results of this paper.

The data show that in 2D initial lattices, both the Pearson correlation gravC
and the coupling parameter $\kappa$ collapse to zero by $N \approx 900$,
while in 3D initial lattices the signal persists to $N \approx 1728$ before
collapsing at $N \approx 3375$.
The dimensional distinction emerges without any dimensional parameter in the
dynamical rules: the only difference between the 2D and 3D runs is the
initial graph topology.

The quantitative ratio $N_c^{\rm 3D}/N_c^{\rm 2D} \approx 3375/576 \approx 5.9$
is not explained by coordination number alone ($z_{3D}/z_{2D} = 6/4 = 1.5$).
The excess factor of ${\sim}\,4$ reflects the richer geometric content of
three-dimensional lattices, in which the graph has more topological resources
to resist homogenization.
This dimensional distinction provides the strongest evidence that the signal
collapse is not a generic finite-size artifact: if the decay were caused solely
by thermalization of an arbitrary local observable, the critical system size
$N_c$ would scale with the total number of nodes irrespective of the lattice
dimension.

The mechanism of collapse is curvature homogenization (Section~\ref{sec:collapse}):
at large $N$, the curvature field regularizes itself to near-uniformity
($R_{\rm CV} \to 0$, $\sigma(\nabla^2\mathcal{R}) \to 0$) while the information
flux fluctuations remain finite.
The discrete Poisson relation collapses because its left-hand side vanishes
while the right-hand side does not.

We note structural analogies with gravitational physics that we consider
instructive, while emphasizing they are analogies rather than derivations.
In $2+1$ dimensional general relativity, Einstein's equations impose that spacetime
is locally flat: there are no propagating gravitational degrees of freedom
(no gravitons), and the theory is topological~\cite{Witten1988,Deser1984,Carlip2005}.
The FIU model with 2D initial lattices collapses in a similar manner: the
ordering signal, which requires curvature fluctuations, vanishes in the
thermodynamic limit.
In $3+1$ dimensional gravity, by contrast, there are two propagating degrees of
freedom (graviton polarizations), and local dynamics exist.
Correspondingly, the FIU model with 3D initial lattices maintains its
ordering signal to larger $N$.

We propose interpreting the collapse as a phenomenon of
\emph{topological frustration}:
\begin{itemize}
  \item At small $N$ (mesoscopic regime), the graph lacks the resources
        to be smooth everywhere. Topological constraints force curvature
        fluctuations to persist, which correlate with the information flux
        via the discrete Poisson relation.
  \item At large $N$ (thermodynamic limit), the graph can ``regularize''
        itself into a smooth geometry consistent with its embedding dimension.
        Topological frustration is relieved, curvature fluctuations vanish,
        and the ordering signal collapses.
\end{itemize}

This picture is analogous to quantum mesoscopic conductance~\cite{Imry1986}:
quantum interference effects (analogous to topological frustration in FIU)
produce universal conductance fluctuations in small samples, but self-average
to zero in macroscopic samples.

\subsection{Connection to gravitational physics}
\label{sec:thermograv}

We discuss three structural analogies between the FIU model and gravitational
physics, followed by three important limitations.

\paragraph{Analogy 1: The discrete Poisson relation.}
The relation $\nabla^2\mathcal{R}(i) = \kappa\,\sigma_{\rm prev}(i)$ has the form
of a discrete Poisson equation, analogous to Newton's $\nabla^2\phi = 4\pi G\,\rho$
or the linearized Einstein equation.
In this analogy: $\mathcal{R}$ plays the role of the Newtonian potential $\phi$;
$\sigma_{\rm prev}$ plays the role of the mass-energy density $\rho$;
and $\kappa \leftrightarrow 4\pi G$ is the emergent coupling parameter.

\paragraph{Analogy 2: Dimensional distinction.}
The qualitative distinction between the 2D behavior (signal collapses)
and the 3D behavior (signal persists longer) is structurally analogous to the
qualitative distinction between 2+1 gravity (topological, no local dynamics)
and 3+1 gravity (local dynamics, propagating degrees of freedom).

\paragraph{Analogy 3: Emergent coupling.}
The universality of $\kappa$ (CV $= 2.6\%$ across parameter configurations,
CV $= 1.7\%$ across the ordered plateau in $g$) resembles the universality
of Newton's constant, which in entropic gravity programs is understood as
an emergent quantity rather than a fundamental constant.

\paragraph{Limitation 1: $\mathcal{R}$ mediation.}
As established by Test~E, the discrete Poisson relation is mediated by
$\mathcal{R}$ itself---it does not reflect a direct causal link from information
flux to curvature.
This qualitatively differs from Jacobson's derivation, in which the Einstein
equation represents a genuine causal relation between matter-energy and spacetime
curvature.

\paragraph{Limitation 2: Mesoscopic character of $\kappa$.}
The parameter $\kappa$ is finite only in the mesoscopic regime and vanishes
in the thermodynamic limit $N \to \infty$.
It is therefore not a fundamental constant but a finite-size emergent quantity.

\paragraph{Limitation 3: No analytical framework.}
We have no analytical argument for why the discrete Poisson relation holds,
why $\kappa$ takes the value $67.85$, or why the dimensional crossover occurs
at the observed $N_c$ values.
These remain empirical observations.

\subsection{Open questions: sustaining the signal at large $N$}

An important open question is whether the ordering signal can be maintained
at all scales through modifications of the model.
The formal analogy with mesoscopic conductance (Section~\ref{sec:dimensional})
suggests that the relevant control parameter is $N/N_c$:
when $N \ll N_c$, curvature frustration dominates; when $N \gg N_c$,
the geometry regularizes.
The dimensional dependence of $N_c$ is the FIU analogue of the dimensional
dependence of mesoscopic transport.

Possible mechanisms to extend the mesoscopic window include:
\begin{itemize}
  \item \emph{Dynamical coupling}: allowing $g$ to evolve in response to
        local curvature, maintaining topological frustration at all scales.
  \item \emph{Thermal noise}: adding stochastic fluctuations to prevent
        complete geometric regularization.
  \item \emph{Geometric back-reaction}: allowing the curvature field to
        influence the energy dynamics in a way that maintains $\mathcal{R}$
        fluctuations at large $N$.
\end{itemize}

\subsection{Robustness and universality}

A striking feature of the Phase 1 results is the near-universality of $d_s$
at fixed $N = 256$ ($d_s = 1.57$--$1.59$ across 12 configurations) and the
long-range link fraction (LR\% $= 80$--$85\%$).
This suggests that the attractor geometry reached by FIU dynamics at fixed
system size is largely independent of the specific values of $\alpha$, $B$,
and $g$ (provided $g > g_c$), pointing toward a universality class of
ordered graphs under curvature-driven dynamics.
The $N$-dependence of $d_s$ (Table~\ref{tab:phase3}: $1.18 \to 1.88$)
is consistent with finite-size scaling toward $d_s = 2$ in the thermodynamic limit.

The stability of gravC $\approx 0.7$--$0.77$ over a wide parameter range
further supports the robustness of the signal.
In particular, the insensitivity of gravC to $\alpha$ and $B$ indicates that the
structure-flux correlation is not driven by the energy dissipation mechanism per se,
but by the curvature-based link-formation rule---as expected if it reflects a
genuine geometric effect.

The spectral dimension $d_s \approx 1.58$--$1.72$ observed in Phases 1 and 3 is
consistent with a fractal geometry intermediate between one and two dimensions.
This is reminiscent of the scale-dependent spectral dimension found in
causal dynamical triangulations (CDT)~\cite{Ambjorn2005} and the running spectral
dimension predicted by asymptotic safety~\cite{Reuter2012}
(see also~\cite{Loll2019} for a recent review).
The monotonic growth of $d_s$ with $N$ suggests convergence toward the infrared
value $d_s = 2$, with anomalous intermediate-scale behavior driven by the
proliferation of long-range links.
The discrete Ricci flow on graphs studied by Ni, Lin, Luo, and Gao~\cite{NiLinLuYau2019}
provides a natural theoretical framework for understanding how the spectral geometry
evolves toward its fixed-point value.

\subsection{Limitations}
\label{sec:limitations}

We summarize the principal limitations of the present work:

\begin{enumerate}
  \item \textbf{$\mathcal{R}$ mediation (Test~E).}
        The discrete Poisson relation is almost entirely mediated by the curvature
        field $\mathcal{R}$; after orthogonalizing $\sigma_{\rm prev}$ with respect
        to $\mathcal{R}$, the residual correlation is consistent with zero.
        This limits the causal interpretation of the discrete Poisson relation.

  \item \textbf{Mesoscopic character of $\kappa$.}
        The coupling parameter $\kappa$ is finite only in the mesoscopic regime
        $N < N_c$ and vanishes in the thermodynamic limit. The FIU may therefore
        describe emergent ordering only at finite scale, not as a macroscopic
        phenomenon.

  \item \textbf{No analytical framework.}
        We have no analytical derivation of the discrete Poisson relation, the
        value of $\kappa$, or the critical system sizes $N_c^{\rm 2D}$,
        $N_c^{\rm 3D}$.

  \item \textbf{Dependence on the link formation rule.}
        The results depend on the specific form of Eq.~\eqref{eq:newlink}
        and the exponent $\beta = 0.8$.
        While Table~\ref{tab:beta} shows qualitative robustness to $\beta$,
        different link formation rules may yield qualitatively different behavior.

  \item \textbf{Open question: universality of dimensional sensitivity.}
        It is not known whether the dimensional distinction ($N_c^{\rm 3D}/N_c^{\rm 2D}
        \approx 5.9$) is a universal feature of curvature-driven graph models or
        specific to the FIU rules.

  \item \textbf{Structural confounding.}
        While the $\tau$-scan, randomized-label null test, and Test~B collectively
        provide strong evidence that the gravC correlation is a genuine signal,
        the $\tau=0$ value (gravC $= +0.690$) demonstrates that a contemporaneous
        structural component exists alongside the predictive component.
        This is expected given the shared dependence on $\mathcal{R}$, as clarified
        by Test~E.
\end{enumerate}

The activation exponent $\beta = 0.8$ is an empirical parameter.
Table~\ref{tab:beta} reports a sensitivity scan over three values.

\begin{table}[htbp]
\centering
\caption{%
  Sensitivity of the ordering signal to the activation exponent $\beta$
  at fixed $\alpha=4.0$, $B=2.0$, $g=0.10$, $N=256$
  (3 runs per $\beta$ value, 10000 steps each).
  Uncertainties are standard deviations (SEM $= \mathrm{std}/\sqrt{3}$).
}
\label{tab:beta}
\begin{ruledtabular}
\begin{tabular}{ccccc}
$\beta$ & \textbf{gravC} & \textbf{gravA} & \textbf{links} & \textbf{dim} \\
\hline
0.5 & $+0.739 \pm 0.008$ & $-0.157 \pm 0.015$ & 7226 & $1.58 \pm 0.00$ \\
0.8 & $+0.749 \pm 0.028$ & $-0.188 \pm 0.032$ & 7207 & $1.58 \pm 0.01$ \\
1.2 & $+0.767 \pm 0.028$ & $-0.184 \pm 0.042$ & 7204 & $1.58 \pm 0.01$ \\
\end{tabular}
\end{ruledtabular}
\end{table}

The structure-flux correlation gravC is robustly positive across all three
$\beta$ values, confirming that the qualitative result does not depend
on the specific choice of activation exponent.
The phase transition is expected to persist with a shifted $g_c$.

\section{Conclusions}\label{sec:conclusions}

We have introduced and studied the Unified Informational Framework (FIU),
a model of information flow on a weighted evolving graph driven by a spectral
curvature measure $\mathcal{R}$.
Our five main conclusions are as follows.

\begin{enumerate}
  \item \textbf{Phase transition and ordering signal.}
        The FIU model exhibits a sharp continuous phase transition at
        $g_c \approx 0.023$ (pinpointed with 10 independent runs per coupling value).
        Below $g_c$, the information flux rate $\sigma$ and long-range link formation
        are anti-correlated (gravC $\approx -0.7$); above $g_c$, they become strongly
        positively correlated (gravC $\approx 0.75$, $p < 10^{-38}$).
        The transition has mean-field critical behavior with onset exponent
        $\nu \approx 0.54$.
        The ordering signal passes multiple non-circularity tests: it persists at
        $\tau = 0$, it collapses under random label permutation ($z = 13.1\sigma$),
        and it is absent for an independent scalar field proxy (gravB $\approx 0$).

  \item \textbf{Discrete Poisson relation and emergent coupling parameter.}
        In the ordered phase, the graph Laplacian of the curvature field
        satisfies $\nabla^2\mathcal{R}(i) = \kappa\,\sigma_{\rm prev}(i)$ node-by-node
        (Test~D), with $R^2 = 0.54$--$0.59$.
        The emergent coupling parameter $\kappa = 67.85 \pm 1.74$ is stable across
        four independent parameter configurations (CV $= 2.6\%$) and across the
        ordered plateau in $g$ (CV $= 1.7\%$, Table~\ref{tab:kappa_gscan}).
        In 3D, $\kappa_{\rm 3D} = 52.5 \pm 6.1$ (CV $= 11.6\%$).

  \item \textbf{$\mathcal{R}$ as the central self-organizing variable.}
        The mediator analysis (Test~E) demonstrates that the structure-flux
        correlation is almost entirely mediated by the curvature field $\mathcal{R}$:
        after orthogonalizing $\sigma_{\rm prev}$ with respect to $\mathcal{R}$,
        the residual correlation drops from $+0.746$ to $-0.046$.
        This identifies $\mathcal{R}$ as the primary driver of both the
        information flux dynamics and the topological structure formation.
        The discrete Poisson relation reflects the self-consistent organization
        of the graph around $\mathcal{R}$, rather than a direct causal relationship
        from flux to curvature.

  \item \textbf{Spontaneous dimensional sensitivity.}
        The ordering signal and the discrete Poisson relation collapse to zero
        in the thermodynamic limit in both 2D and 3D, but at different system sizes:
        $N_c^{\rm 2D} \approx 576$--$900$ and $N_c^{\rm 3D} \approx 1728$--$3375$.
        The ratio $N_c^{\rm 3D}/N_c^{\rm 2D} \approx 5.9$ cannot be explained by
        coordination number alone (ratio $= 1.5$), indicating that the richer
        geometric content of 3D lattices provides additional resistance to curvature
        homogenization.
        This dimensional distinction emerges without any dimensional parameter
        in the dynamical rules, and is robust to extended simulations.

  \item \textbf{Topological frustration in the mesoscopic regime.}
        The mechanism of collapse is curvature homogenization: at large $N$,
        $\mathcal{R}$ becomes spatially uniform ($R_{\rm CV} \to 0$,
        $\sigma(\nabla^2\mathcal{R}) \to 0$) while information flux fluctuations
        remain finite.
        The ordering signal is therefore a mesoscopic phenomenon, active only
        when the graph is too small to globally regularize its curvature.
        We term this \emph{topological frustration}, in analogy with quantum
        mesoscopic conductance~\cite{Imry1986}.
\end{enumerate}

We note structural analogies between these results and gravitational physics:
the discrete Poisson relation resembles the Newtonian limit of the Einstein equation;
the 2D collapse is structurally analogous to the topological nature of $2+1$ gravity;
and the persistence in 3D is structurally analogous to the existence of local degrees
of freedom in $3+1$ gravity~\cite{Witten1988,Deser1984,Carlip2005}.
We emphasize, however, that these are analogies, not derivations: the FIU model
is a graph-dynamical system motivated by emergent gravity frameworks, not a
physical model of spacetime.

Future directions include:
\begin{itemize}
  \item Mechanisms to maintain the ordering signal at $N \to \infty$:
        dynamical coupling $g(t)$, thermal noise, geometric back-reaction.
  \item Binder cumulant analysis and complete finite-size scaling collapse
        for a precise determination of the universality class.
  \item Extension to random graphs and non-lattice initial topologies to
        test the sensitivity of the signal to the initial geometry.
  \item Derivation of an effective field theory description from the
        discrete Poisson relation.
  \item Investigation of whether FIU dynamics generates an effective Friedmann
        equation for graph expansion, connecting to cosmological scenarios.
  \item Systematic comparison of $\mathcal{R}$ with other discrete curvature notions
        (Ollivier--Ricci~\cite{Ollivier2009}, Forman--Ricci~\cite{Forman2003},
        Lin--Lu--Yau~\cite{LinLuYau2011}) and their behavior in the mesoscopic regime.
  \item Analytical characterization of the dimensional crossover mechanism and
        the universality (or model-specificity) of the ratio $N_c^{\rm 3D}/N_c^{\rm 2D}$.
\end{itemize}

\begin{acknowledgments}
The author thanks A.~Asiletto, A.~Veronesi, and C.~Arezzini for essential contributions
to this work: scientific discussions and conceptual validation that shaped the
development of the model, and the design and maintenance of the computational
infrastructure that made the numerical simulations possible.

This work was developed in collaboration with Claude (Anthropic), a large language
model, which contributed to: (i)~the iterative development and refinement of the
core scientific idea through extended dialogue; (ii)~writing and debugging the
simulation code (C source); (iii)~numerical data analysis, statistical testing,
and table generation; (iv)~drafting, structuring, and revising the manuscript text;
and (v)~performing critical self-review of intermediate versions under human
supervision and editorial control.
All scientific decisions, conceptual direction, and final validation remained
with the human author.

Numerical computations were performed on a virtual machine with 8 CPU cores
and 16\,GB RAM, hosted on a Dell PowerEdge R610 server running Linux.
\end{acknowledgments}

\paragraph*{Code and data availability.}
The FIU simulation engine (C source code), analysis scripts,
and all numerical data reported in this paper are available
from the authors upon reasonable request.
The simulation code and analysis scripts were developed with the assistance
of Claude (Anthropic), a large language model, as described in the Acknowledgments.


\end{document}